# Time Domain Differential Equation Based Fault Location Identification in Mixed Overhead-Underground Power Distribution Systems

Ali Shakeri Kahnamouei, *Student Member, IEEE*, and Saeed Lotfifard, *Senior Member, IEEE*

*Abstract*—**This paper proposes a time-domain fault location identification method for mixed overhead-underground power distribution systems that can handle challenging fault scenarios such as sub-cycle faults, arcing faults and evolving faults. The proposed method is formulated based on differential equations of the system and accounts for the peculiarities of power distribution systems with distributed generations. It considers the impacts of loads, multi-phase laterals and sub-laterals, heterogenous overhead and underground lines, and infeeds and remote-end fault current contributions of distributed generations. It utilizes data collected by limited number of measuring devices installed in modern power distribution systems to systematically eliminate possible multiple fault location estimations to provide a single correct estimation of the actual location of the fault. The performance of the proposed method is demonstrated using IEEE 34-node test system.**

## I. INTRODUCTION

FAULT location identification programs are crucial functions for enabling fast and automated restoration of faulted distribution systems, leading to enhanced fault resilience and reliability of the system [1]-[6]. Due to their importance, a variety of methods have been developed in the past, such as impedance-based methods [7], voltage sag data-based methods [8], traveling-based methods [9], and phasor measurement unit (PMU)-data-based methods [10]. The strengths and weaknesses of the methods have been investigated in several studies, such as [11]-[15].

Phasor-domain fault location methods assume the fault duration is long enough to capture the required during-fault data to extract accurate phasors from the collected data. Although this assumption is correct in most fault cases, there are challenging fault scenarios in which the during-fault data set is too small to extract reliable voltage and current phasors for the fault location applications. For instance, in the case of sub-cycle or intermittent faults, the fault duration could be as short as ¼ of a cycle [16]. In [17], some real-world scenarios of such cases due to the presence of underground laterals in distribution systems are presented. Another example case of faults with limited during-fault data is fast-evolving faults in which the fault starts with a given fault type and evolves to another fault type. If such a change in the fault types occurs in a short period of time, enough data may not be available during each type of fault for accurate phasor-based fault location identification purpose.

Time-domain fault location methods are promising solutions that can operate properly even in the case of faults with such short durations, as they only require a few samples of data to estimate the location of the fault. Common time-domain fault location methods are traveling-wave methods and differential equation-based algorithms. Traveling wave methods have been implemented for transmission line protection and fault analysis. An example of early practical implementation and testing of such methods is described in [18]. Different methods are also proposed to utilize traveling wave methods for fault location in distribution systems [19]-[21]. The presence of short lines, lateral and sub-laterals, and the need for high sampling rate data collection and communication are some of the challenges for implementing traveling-wave methods for fault location identification in distribution systems.

Differential equation-based methods have also been developed for transmission line protection [22]. Efforts have been made to implement such differential equation-based protection methods for distribution systems fault location applications [23]-[27]. The above studies have demonstrated the effectiveness of differential equation-based methods for distribution systems fault location applications. However, they are developed for conventional power distribution systems and do not consider the presence of distributed generations (DGs). When DGs are installed in distributed systems, they create remote-end and infeed current contributions to the fault, which should be considered in the process of identifying the fault location. Moreover, in some of the methods, the possible faults on laterals are ignored, and only faults on the main feeder are considered. Unlike transmission lines, in distribution systems, a limited number of measuring devices are installed in the system. Therefore, between the faulted point and the measuring devices, several loads, laterals, and DGs may exist, and their impacts should be considered. In fault scenarios that fault current is not significantly higher than the load currents, impacts of loads and laterals currents become more influential on the fault location results. Examples of such scenarios are (a) when the short circuit capacity of the upstream transmission line is not significantly high, which is the case in transmission systems with high penetration of inverter-based renewable sources, (b) heavily loaded distribution systems with large loads, (c) and faults with large fault resistance.

A. Shakeri Kahnamouei, and S. Lotfifard are with the School of Electrical Engineering and Computer Science, Washington State University, Pullman, WA, 99164, USA, (e-mail: a.shakerikahnamouei@wsu.edu, and s.lotfifard@wsu.edu).

To address the above challenges, this paper proposes a time domain differential equation-based fault location method that explicitly takes into account the peculiarities of distribution systems with distributed generations. Specifically, the contributions of the paper are as follows:

1) A time domain fault location method is proposed that remote-end and infeed current contribution to faults are precisely taken into account. This is a crucial consideration for accurate fault location identification in distribution systems with the presence of DGs.

2) The presence of laterals, sub-laterals, and loads is explicitly considered in the proposed method, which enhances the accuracy of the results. A time-domain backward/forward sweep (BFS) method is implemented that precisely calculates the impacts of laterals, loads, and remote end currents on the fault current. This feature is crucial for enhancing the accuracy of the fault location identification results, especially in systems with large loads and/or systems with low short circuit capacities.

3) The fault impedance is not ignored, and arcing faults are considered.

4) Due to the presence of laterals and sub-laterals, multiple locations may exist that are at the same electrical distances from the substation. A systematic approach is implemented to eliminate such multiple fault location identifications, providing a single estimation of the actual location of fault.

The remainder of this paper is organized as follows: Section II presents the proposed fault location method. The case study results are presented in section III. Finally, the conclusions are presented in section IV.

## II. PROPOSED FAULT LOCATION IDENTIFICATION METHOD

A typical faulted distribution system is shown in Fig.1-(a), which represents IEEE 34 node distribution system. Fig. 1-(b) illustrates the $\pi$ model of the faulted line between nodes 828 and 830 where $k = 828$ and $k + 1 = 830$ are the upstream and downstream nodes of the faulted line, respectively. R, L, and C denote the line resistance, inductance, and shunt capacitance, respectively. The following subsections explain the proposed time-domain fault location method.

### A. Fault Location Formulation

As indicated before, unlike transmission lines, in distribution systems, only a limited number of measurements are available. Therefore, it is not realistic to assume in Fig.1-(b) that measurements are available from both ends of the faulted line. For ease of explaining the proposed method, in this subsection, the proposed fault location method is explained assuming values of $v_k$, $v_{k+1}$, $i_k$, $i_{k+1}$ and $i_F$ in Fig. 1-(b) are known. In Subsection II-B, the proposed method for calculating these values from measured data from a few measuring devices installed in the system is explained.

It should also be mentioned that clearly, the actual faulted line is unknown before the fault location identification is completed. Therefore, each line of the system is considered a potential faulted line. Starting from the substation node, each line of the distribution system is checked if it is the faulted line.

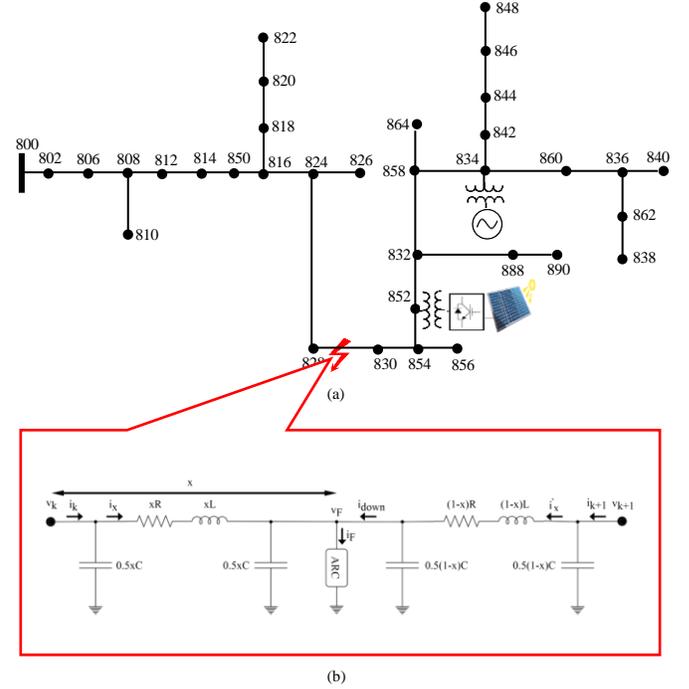

Fig. 1. (a) Schematic of a typical faulted distribution system (b) $\pi$ model of the faulted line between node k=828 and k+1= 830

In Fig.1-(b), $x$ denotes the fault distance in the selected line, which is estimated using the procedures explained in section II-A-4. If the estimated value of $x$ for a given line is $0 \leq x \leq 1$, that line enters into the short list of the potential faulted lines. After applying the algorithm to all lines of the system one at a time, if there is only one line in the short list of faulted lines, that line is the faulted line; otherwise, the actual faulted line is determined using the procedure explained in Section II-A-4.

Note that in the fault location formulation development, the term "faulted line" refers to the line which is selected as a potential faulted line. As explained above, whether it is the actual faulted line or not will be determined after applying the proposed FL algorithm.

It should also be mentioned that unlike protection functions which are online functions, fault location functions are off-line functions. This means unlike protection algorithms that should be able to make decisions in real-time as new samples of data become available, fault location methods are used after the operation of protection functions to identify the actual location of the fault. Therefore, fault location algorithms can have access to the whole during-fault data as well as pre-fault data. This provides significant advantages, such as the ability to design more robust algorithms by utilizing whole during fault data rather than using a fraction of data and the ability to use advanced signal processing methods for denoising the collected data.

### 1) Single-Phase to Ground Faults

In this subsection, to explain the fault location method, a single-phase to ground fault, A-G, is considered and in Subsection II-A-2 it is explained how the method can be applied to other types of faults.

In general, differential equation-based fault location methods are developed based on Kirchhoff's voltage law

(KVL), and Kirchhoff's current law (KCL) applied to the faulted line (i.e., Fig.1-(b)). This procedure has been used in [23], [27]. This paper also follows a similar approach as [23] for formulating the problem of estimating the fault distance within the faulted line (i.e., $x$ in Fig.1-(b)). However, in contrast to the above-mentioned papers in which the problem of finding x is formulated based on $v_k$, $i_k$, $i_{k+1}$, this paper formulates the problem based on $v_k$, $i_k$, $i_{k+1}$, and $v_{k+1}$. Utilizing $v_{k+1}$ in the estimation procedure enables more accurate downstream system representation (which will be discussed in Section II-C) in the fault location estimation process. Moreover, in this paper, procedures are developed to implement the fault location method for multi-phase fault cases. In Fig. 1-(b), the following equations can be written at each time instant:

$$v_k = x\left(Ri_x + L\frac{di_x}{dt}\right) + v_F \tag{1}$$

$$v_{k+1} = (1-x)\left(Ri'_x + L\frac{di'_x}{dt}\right) + v_F \tag{2}$$

Where

$$i_x = i_k - \frac{1}{2}xC\frac{dv_k}{dt} \tag{3}$$

$$i'_x = i_{k+1} - \frac{1}{2}(1-x)C\frac{dv_{k+1}}{dt} \tag{4}$$

Note that $R$, $L$, and $C$ are 3×3 matrices representing the line resistance, inductance, and capacitance, respectively. Therefore, the mutual couplings of the lines are also considered.

Different arc fault models have been used for the fault location application. For instance, in [26] single parameter arc model presented in [28] is used. In [29] three parameter model is utilized, while [23] uses four parameter model proposed by [30]. In this paper the four parameter model is used as follows [23]:

$$v_F = R_F i_F + L_F\frac{di_F}{dt} + v_{Fp}sg^+(i_F) + v_{Fn}sg^-(i_F) \tag{5}$$

Where $R_F$ and $L_F$ denote fault resistance and inductance, respectively. The arc voltage is positive and negative in the positive and negative semi-cycles of the arc fault, respectively. $v_{Fp}$ and $v_{Fn}$ represent the positive and negative arc voltages, respectively. The positive and negative arc voltages cannot be present at the same time, and they are turned on and off as follows:

$$sg^+(i_F) = \begin{cases} 1, & i_F > 0 \\ 0, & i_F \le 0 \end{cases} \tag{6}$$

$$sg^-(i_F) = \begin{cases} 0, & i_F \ge 0 \\ -1, & i_F < 0 \end{cases} \tag{7}$$

By replacing (3)-(4), and (5) into (1)-(2), the following equations can be derived:

$$v_k = xv_z + x^2v_y + R_Fi_F + L_F\frac{di_F}{dt} + v_{Fp}sg^+(i_F) + v_{Fn}sg^-(i_F) \tag{8}$$

$$v_{k+1} = (1-x)v'_z + (1-x)^2v'_y + R_Fi_F + L_F\frac{di_F}{dt} + v_{Fp}sg^+(i_F) + v_{Fn}sg^-(i_F) \tag{9}$$

$$v_z = Ri_k + L\frac{di_k}{dt} \tag{10}$$

$$v_y = -\frac{1}{2}C\left(R\frac{dv_k}{dt} + L\frac{d^2v_k}{dt^2}\right) \tag{11}$$

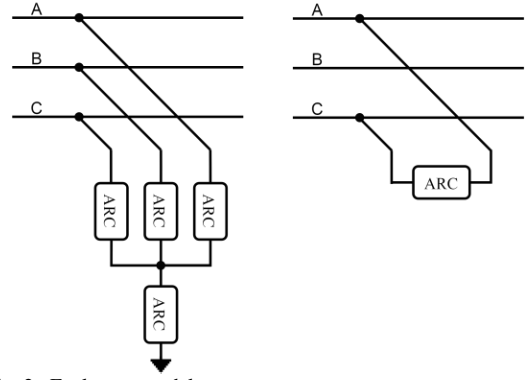

Fig. 2. Fault type models

$$v'_z = Ri_{k+1} + L\frac{di_{k+1}}{dt} \tag{12}$$

$$v'_y = -\frac{1}{2}C\left(R\frac{dv_{k+1}}{dt} + L\frac{d^2v_{k+1}}{dt^2}\right) \tag{13}$$

By neglecting the shunt capacitors of the faulted line in Fig.1-(b), terms $v_y$ and $v'_y$ are removed from (8) and (9), respectively. Therefore, (8)-(9) become linear. Ignoring the shunt capacitors of the faulted line is a reasonable assumption as the currents of the shunt capacitors in comparison to the fault current are negligible.

Equations (8)-(9) can be rewritten in a matrix form for A-G fault as follows:

$$\begin{bmatrix} v_k^a \\ v_{k+1}^a - v_z'^a \end{bmatrix}$$
$$= \begin{bmatrix} v_z^a & i_F^a & \frac{di_F^a}{dt} & sg^+(i_F^a) & sg^-(i_F^a) \\ -v_z'^a & i_F^a & \frac{di_F^a}{dt} & sg^+(i_F^a) & sg^-(i_F^a) \end{bmatrix} \begin{bmatrix} x \\ R_F^a \\ L_F^a \\ v_{Fp}^a \\ v_{Fn}^a \end{bmatrix} \tag{14}$$

Where

$$v_k^a = \begin{bmatrix} v_{k_1}^a & v_{k_2}^a & \cdots & v_{k_{N_F}}^a \end{bmatrix}^T \tag{15}$$

$$v_{k+1}^a = \begin{bmatrix} v_{k+1_1}^a & v_{k+1_2}^a & \cdots & v_{k+1_{N_F}}^a \end{bmatrix}^T \tag{16}$$

$$v_z^a = \begin{bmatrix} v_{z_1}^a & v_{z_2}^a & \cdots & v_{z_{N_F}}^a \end{bmatrix}^T \tag{17}$$

$$v_z'^a = \begin{bmatrix} v_{z_1}'^a & v_{z_2}'^a & \cdots & v_{z_{N_F}}'^a \end{bmatrix}^T \tag{18}$$

$$i_F^a = \begin{bmatrix} i_{F_1}^a & i_{F_2}^a & \cdots & i_{F_{N_F}}^a \end{bmatrix}^T \tag{19}$$

$$\frac{di_F^a}{dt} = \begin{bmatrix} \frac{di_{F_1}^a}{dt} & \frac{di_{F_2}^a}{dt} & \cdots & \frac{di_{F_{N_F}}^a}{dt} \end{bmatrix}^T \tag{20}$$

$$sg^+(i_F^a) = \begin{bmatrix} sg^+(i_{F_1}^a) & sg^+(i_{F_2}^a) & \cdots & sg^+\left(i_{F_{N_F}}^a\right) \end{bmatrix}^T \tag{21}$$

$$sg^-(i_F^a) = \begin{bmatrix} sg^-(i_{F_1}^a) & sg^-(i_{F_2}^a) & \cdots & sg^-\left(i_{F_{N_F}}^a\right) \end{bmatrix}^T \tag{22}$$

### 2) Multi-Phase to Ground Faults

Equation (14) can be extended to multi-phase to ground faults. In Fig. 2, assume an A-B-G fault. According to KVL, the fault voltage can be written for phases A and B as follows:

$$v_F^a = R_F^a i_F^a + L_F^a\frac{di_F^a}{dt} + v_{Fp}^a sg^+(i_F^a) + v_{Fn}^a sg^-(i_F^a) + v_{Fg} \tag{23}$$

$$v_F^b = R_F^b i_F^b + L_F^b\frac{di_F^b}{dt} + v_{Fp}^b sg^+(i_F^b) + v_{Fn}^b sg^-(i_F^b) + v_{Fg} \tag{24}$$

Where

$$v_{Fg} = R_{Fg}i_{Fg} + L_{Fg}\frac{di_{Fg}}{dt} + v_{Fp_g}sg^+(i_{Fg}) + v_{Fn_g}sg^-(i_{Fg}) \quad (25)$$

$$i_{Fg} = i_F^a + i_F^b \quad (26)$$

Equation (14) can be expanded for A-B-G faults, as shown at the bottom of the page in (27).

### 3) Phase to Phase Faults

To develop the FL equation for the phase-to-phase faults, assume an A-B fault in Fig. 2. The KVL constraint at the faulted point is as follows:

$$v_F^a = R_F i_F^a + L_F \frac{di_F^a}{dt} + v_{Fp}sg^+(i_F^a) + v_{Fn}sg^-(i_F^a) + v_F^b \quad (28)$$

Therefore, (14) can be expanded for A-B faults as follows:

$$\begin{bmatrix} v_k^a - v_k^b \\ (v_{k+1}^a - v_z'^a) - (v_{k+1}^b - v_z'^b) \end{bmatrix}$$
$$= \begin{bmatrix} v_z^a - v_z^b & i_F^a & \frac{di_F^a}{dt} & sg^+(i_F^a) & sg^-(i_F^a) \\ -(v_z'^a - v_z'^b) & i_F^a & \frac{di_F^a}{dt} & sg^+(i_F^a) & sg^-(i_F^a) \end{bmatrix} \begin{bmatrix} x \\ R_F \\ L_F \\ v_{Fp} \\ v_{Fn} \end{bmatrix} \quad (29)$$

### 4) Estimating Fault Parameters using Weighted Least Square Estimation

Equation (14) is a set of linear equations that can be solved by so-called weighted least square state estimation [31] and is represented as follows:

$$Z = X\theta + \epsilon \quad (30)$$

Where

$$Z = \begin{bmatrix} v_k^a \\ v_{k+1}^a - v_z'^a \end{bmatrix} \quad (31)$$

$$X = \begin{bmatrix} v_z^a & i_F^a & \frac{di_F^a}{dt} & sg^+(i_F^a) & sg^-(i_F^a) \\ -v_z'^a & i_F^a & \frac{di_F^a}{dt} & sg^+(i_F^a) & sg^-(i_F^a) \end{bmatrix} \quad (32)$$

$$\theta = [x \quad R_F^a \quad L_F^a \quad v_{Fp}^a \quad v_{Fn}^a]^T \quad (33)$$

Where $Z$ is the measured vector of samples, $X$ is a known matrix which is called regressors matrix, and $\theta$ is the estimated vector of parameters. Also, $\epsilon$ denotes the error vector.

The classical WLSE approach to estimate the parameters in (30) is to minimize the quadratic norm of the error vector. Consequently, the final solution is as follows:

$$\theta = (X^T W X)^{-1} X^T W Z \quad (34)$$

Where $W$ is the weighting vector of the data samples.

To enhance the accuracy of the results, the following function is proposed for determining the weighting factors.

$$W_n^m = \omega^m \left(1 - e^{-\frac{n}{0.1N_F}}\right) \quad (35)$$

Where $N_F$ denotes the total number of collected samples during the fault duration and $\omega^m$ represents the weighting factor of $m^{th}$ measurement. Also, $W_n^m$ shows the weighting factor of the $n^{th}$ sample of the $m^{th}$ measurement. The above function assures samples that are taken just after the occurrence of the fault, during which the transient fluctuation of the fault current is higher, have lesser impacts on the estimation results.

### B. Estimating Values of $i_k$ and $v_k$

As indicated before, only a limited number of measuring devices are installed in distribution systems. Therefore, the values of $i_k$ and $v_k$ (i.e., upstream voltage and current) of the faulted line in Fig.1-(b) should be estimated using available measuring devices located at other nodes of the system.

The substation is located at the upstream side of the faulted line, and voltage and current measurements are available at the substation. If other measuring devices collecting synchronized samples between the substation and node $k$ exist, which can provide sampled values of both voltage and current signals, the closest measuring device to node $k$ is set as the root node. Note that for a measuring device to be considered as a root node, it should be able to provide sampled values of both voltage and current signals. A measuring device that only measures voltage or current signals cannot be considered a root node. Such data can be used for determining the actual location of the fault, which will be discussed in Section II-D.

Starting from the root node, voltages of all nodes and line currents can be easily calculated using KVL and KCL if loads or laterals do not exist. For instance, Fig. 3 shows a portion of a distribution system. Assume $v_j$ and $i_j$ are measured by a measuring device (i.e., node n is the root node). $v_{j+1}$ can be calculated as follows:

$$v_{j+1} = v_j - R\left(i_j - 0.5c\frac{dv_j}{dt}\right) - L\frac{d(i_j - 0.5c\frac{dv_j}{dt})}{dt} \quad (36)$$

Then, $i_{j+1}$ can be calculated as follows:

$$\begin{bmatrix} v_k^a \\ v_{k+1}^a - v_z'^a \\ v_k^b \\ v_{k+1}^b - v_z'^b \end{bmatrix} = \begin{bmatrix} v_z^a & i_F^a & 0 & i_{Fg} & \frac{di_F^a}{dt} & 0 & \frac{di_{Fg}}{dt} & sg^+(i_F^a) & sg^-(i_F^a) & 0 & 0 & sg^+(i_{Fg}) & sg^-(i_{Fg}) \\ -v_z'^a & i_F^a & 0 & i_{Fg} & \frac{di_F^a}{dt} & 0 & \frac{di_{Fg}}{dt} & sg^+(i_F^a) & sg^-(i_F^a) & 0 & 0 & sg^+(i_{Fg}) & sg^-(i_{Fg}) \\ v_z^b & 0 & i_F^b & i_{Fg} & 0 & \frac{di_F^b}{dt} & \frac{di_{Fg}}{dt} & 0 & 0 & sg^+(i_F^b) & sg^-(i_F^b) & sg^+(i_{Fg}) & sg^-(i_{Fg}) \\ -v_z'^b & 0 & i_F^b & i_{Fg} & 0 & \frac{di_F^b}{dt} & \frac{di_{Fg}}{dt} & 0 & 0 & sg^+(i_F^b) & sg^-(i_F^b) & sg^+(i_{Fg}) & sg^-(i_{Fg}) \end{bmatrix} \begin{bmatrix} x \\ R_F^a \\ R_F^b \\ R_{Fg} \\ L_F^a \\ L_F^b \\ L_{Fg} \\ v_{Fp}^a \\ v_{Fn}^a \\ v_{Fp}^b \\ v_{Fn}^b \\ v_{Fp_g} \\ v_{Fn_g} \end{bmatrix} \quad (27)$$

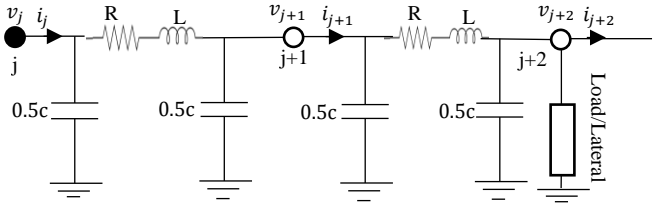

Fig. 3. A portion of a distribution system presented by π model

$$i_{j+1} = i_j - 0.5c\frac{dv_j}{dt} - 0.5c\frac{d(v_{j+1})}{dt} \quad (37)$$

Known values of $v_{j+1}$ and $i_{j+1}$, $v_{j+2}$ can be calculated similar to (36). However, $i_{j+2}$ cannot be calculated similar to (37) as currents of the load/lateral at node $j + 1$ should be calculated. In the following subsections, methods are proposed to calculate the current of loads and laterals.

Note that if a DG is connected to node $j + 1$ in Fig. 3, its presence can be modeled based on the output current of the DG. Therefore, instead of (37), the following equation should be used.

$$i_{j+1} = i_j - 0.5c\frac{dv_j}{dt} - 0.5c\frac{d(v_{j+1})}{dt} + i_{DG} \quad (38)$$

Where $i_{DG}$ is the output current of the DG. If the system only has a few DGs with small capacities, $i_{DG}$ can be ignored. However, for systems with DGs with large capacities, $i_{DG}$, which is measured at the point of coupling of DGs, should be incorporated into the equations to consider the contribution of DGs to faults.

### 1) Time-Domain Load Current Calculation

At a given node of the system (for instance, node $j + 2$ in Fig. 3), once the voltage of that node is determined based on the voltage of the upstream node (i.e., node $j + 1$ in Fig. 3), load currents at that given node (i.e., node $j + 2$ in Fig. 3) are calculated as follows:

Assume the voltages of the load for a specified number of time instants are given. Kirchhoff's voltage law for the load is as follows:

$$v = R_l i + L_l \frac{di}{dt} \quad (39)$$

Where $R_l$ and $L_l$ denote the load's resistance and inductance, respectively. $R_l$ and $L_l$ of the load are calculated as follows:

$$R_l = Real\left(\frac{V^2}{\sqrt{P^2 + Q^2}}\right) \quad (40)$$

$$L_l = \frac{Imag\left(\frac{V^2}{\sqrt{P^2 + Q^2}}\right)}{2\pi f} \quad (41)$$

Where $P$ and $Q$ are active and reactive power of the load and $V$ and $f$ are the base voltage value and frequency of the system, respectively. Note that (40)-(41) model the load as constant impedance, which is commonly used in time domain calculations. However, if voltage dependent load model is preferred, the following approach can be used, which is based on a commonly used approach in time domain simulation software such as PSCAD. Every half cycle (40) and (41) are updated based on the calculated RMS (root mean square) value of the voltage to reflect the dependency of the load as a function of the RMS voltage value of the system. If the fault duration is less than a half cycle, the load value is not updated and is kept as the initial calculated value based on (40)-(41).

The sampling interval of the measurements is defined as

$$t_s = \frac{1}{f_s} \quad (42)$$

Where $f_s$ and $t_s$ are the sampling frequency and sampling interval respectively. The first-order approximation of the first derivative of the current for sample $n$ is defined as follows:

$$\frac{di_n}{dt} = \frac{i_{n+1} - i_n}{t_s} \quad (43)$$

By replacing (43) into (39), equations (44) and (45) can be written for $n$ and $n - 1$ samples, respectively.

$$v_n = R_l i_n + \frac{L_l}{t_s}(i_{n+1} - i_n) \quad (44)$$

$$v_{n-1} = R_l i_{n-1} + \frac{L_l}{t_s}(i_n - i_{n-1}) \quad (45)$$

Therefore, current of the load at $n^{th}$ time sample is derived from (45) as follows:

$$i_n = \frac{t_s}{L_l}v_{n-1} + \left(1 - \frac{t_s R_l}{L_l}\right)i_{n-1} \quad (46)$$

The second statement of (46) indicates that to calculate the current at each time instant, the sample of current in the previous time instant is required, which is not known for the first sample (i.e., the initial point $n = 0$). To initialize (46), the collected pre-fault voltage and current data at the root node are used to perform a phasor-based distribution system power flow analysis to obtain the pre-fault phasors of load currents. The calculated phasor of the pre-fault load current is represented in the time domain, and the required pre-fault current sample (i.e. $i_{n-1}$) is used in (46).

### 2) Time-Domain Backward/Forward Sweep Method

Between the faulted line (i.e., the line between nodes 828 and 830 in Fig.1-(a)) and the root node laterals may also exist (i.e., branches connected to nodes 808, 816, and 824 in Fig.1-(a)). To calculate the current of the lateral, the following time domain backward/forward sweep method is implemented.

For a given lateral, N samples of the voltage of the first node of the lateral (e.g., node 816 in Fig.1-(a)) are calculated by using the measured current and voltages at the root node. It should be noted that the considered samples might include any number of pre-fault and post-fault samples. Therefore, the calculated voltages and currents can start from a pre-fault sample and continue to a post-fault sample, including all fault period samples.

In the first step, the voltages of all nodes of the lateral (i.e., nodes 818, 820, and 822 in Fig.1-(a)) are assumed to be equal to the voltage of the first node of the lateral (i.e., node 816 in Fig.1-(a)) for all N samples. Then, the derivative of the voltage is calculated using the second-order approximation of the first derivative as follows:

$$\frac{dv_n}{dt} = \frac{v_{n+1} - v_{n-1}}{2t_s} \quad (47)$$

After calculating voltage derivatives, the current flowing through the shunt capacitors of the lines at time step $n$ is calculated as follows:

$$\begin{bmatrix} i_{c_n}^a \\ i_{c_n}^b \\ i_{c_n}^c \end{bmatrix} = \frac{1}{2} \begin{bmatrix} C^{aa} & C^{ab} & C^{ac} \\ C^{ba} & C^{bb} & C^{bc} \\ C^{ca} & C^{cb} & C^{cc} \end{bmatrix} \begin{bmatrix} \dfrac{dv_n^a}{dt} \\ \dfrac{dv_n^b}{dt} \\ \dfrac{dv_n^c}{dt} \end{bmatrix} \quad (48)$$

The load currents are also calculated using the proposed method in II-B-1. Then, after calculating load and shunt capacitor currents, the currents flowing through the lines can be calculated starting from the last node of the lateral. Thereafter, the derivatives of the line currents are calculated using a similar equation as (47).

After calculating the line currents and their derivatives for all samples, the node voltages are updated for all samples starting from the first node of the lateral as follows:

$$v_{j_n} = v_{i_n} - \left( R_{ij} i_{ij_n} + L_{ij} \frac{di_{ij_n}}{dt} \right) \quad (49)$$

Where $v_{i_n}$ and $v_{j_n}$ are the voltage values of node $i$ and $j$ at $n^{th}$ time step, respectively. $R_{ij}$ and $L_{ij}$ denote the resistance and inductance of the line between node $i$ and $j$, respectively, and $i_{ij_n}$ represents the line current flowing from node $i$ to $j$ at $n^{th}$ time instant.

Equation (49) can be expanded to three phase form as follows:

$$\begin{bmatrix} v_{j_n}^a \\ v_{j_n}^b \\ v_{j_n}^c \end{bmatrix} = \begin{bmatrix} v_{i_n}^a \\ v_{i_n}^b \\ v_{i_n}^c \end{bmatrix} - \begin{bmatrix} R_{ij}^{aa} & R_{ij}^{ab} & R_{ij}^{ac} \\ R_{ij}^{ba} & R_{ij}^{bb} & R_{ij}^{bc} \\ R_{ij}^{ca} & R_{ij}^{cb} & R_{ij}^{cc} \end{bmatrix} \begin{bmatrix} i_{ij_n}^a \\ i_{ij_n}^b \\ i_{ij_n}^c \end{bmatrix}$$
$$- \begin{bmatrix} L_{ij}^{aa} & L_{ij}^{ab} & L_{ij}^{ac} \\ L_{ij}^{ba} & L_{ij}^{bb} & L_{ij}^{bc} \\ L_{ij}^{ca} & L_{ij}^{cb} & L_{ij}^{cc} \end{bmatrix} \begin{bmatrix} \dfrac{di_{ij_n}^a}{dt} \\ \dfrac{di_{ij_n}^b}{dt} \\ \dfrac{di_{ij_n}^c}{dt} \end{bmatrix} \quad (50)$$

The BFS is an iterative method that stops when the difference between the calculated voltages of all the samples at the present and previous iterations is less than a specified value. The error is calculated using (67), which is defined in section II-D.

To numerically implement the above time-domain BFS using the collected samples of data, the following procedure should be followed. One of the issues that should be considered is tracking the sequence of data samples when derivative terms exist in the equations. For instance, as explained above, at the first iteration, it is assumed that the values of voltages of all nodes of the lateral are equal to the value of the voltage of the first node of the lateral. To calculate shunt capacitor currents in (48), the derivates of the voltage of nodes should be calculated using (47). According to (47), since at each time instant the derivative is calculated using the samples of data at the previous and next time instants, the derivative cannot be calculated at the first and last time instants of the available data set, since there are no samples of data before the first sample and after the last sample of the dataset. Therefore, the shunt capacitors currents in (48) cannot be calculated at the first and last time instants. Thus, the first and last time instants are discarded from the set of the time instants. To keep track of the discarded time instants from the start and end of the set of samples, variables $s$ and $e$ are defined as the number of discarded time instants from the start and end of the samples, respectively.

At the first iteration, $s_v^{iter}$ and $e_v^{iter}$ are zero for the voltage values. For the voltage derivatives, the followings hold:

$$s_{dv} = s_v^{iter} + 1 \quad (51)$$
$$e_{dv} = e_v^{iter} + 1 \quad (52)$$

As the shunt capacitor currents are calculated according to (48), $s_c = s_{dv}$ and $e_c = e_{dv}$.

Load currents are calculated according to (46), in which the load current of the time step after the last sample can also be calculated. Therefore, $s$ and $e$ variables for load currents are as follows:

$$s_{load} = s_v^{iter} \quad (53)$$
$$e_{load} = e_v^{iter} - 1 \quad (54)$$

After calculating all load currents and shunt capacitor currents, the line currents are calculated. Consequently, $s_{line}$ and $e_{line}$ can be updated as follows:

$$s_{line} = max\{s_c, s_{load}, s_{line\_downstream}\} \quad (55)$$
$$e_{line} = max\{e_c, e_{load}, e_{line\_downstream}\} \quad (56)$$

Where $s_{line\_downstream}$ and $e_{line\_downstream}$ in the right-hand-side of (55)-(56) are $s$ and $e$ values of the downstream line. Thereafter, the derivatives of the line currents are calculated using (47). As explained for the voltage derivatives, derivatives at the first and last data samples cannot be calculated here as well. Therefore,

$$s_{dline} = s_{line} + 1 \quad (57)$$
$$e_{dline} = e_{line} + 1 \quad (58)$$

In the forward stage, all bus voltages are calculated using (49). According to (49), $s_v^{iter}$ and $e_v^{iter}$ can be obtained as follows:

$$s_v^{iter+1} = max\{s_v^{iter}, s_{dline}\} \quad (59)$$
$$e_v^{iter+1} = max\{e_v^{iter}, e_{dline}\} \quad (60)$$

It should be mentioned that since at each iteration a few samples are discarded from the start and end of the total number of samples due to the derivative terms, it is recommended to include at least a few samples of pre-fault and post-fault data in the dataset. However, even without including any samples from pre-fault and post-fault periods, the proposed method works. The difference is that in the latter case, a few data samples from the start and end of the during fault samples will be discarded and will not be used in the fault location identification process of Section II-A.

## C. Estimating Values of $i_{k+1}$, $v_{k+1}$, and $i_F$

A common approach in fault location methods for representing the impacts of the downstream networks in the fault location process is lumping up the downstream loads and modeling them as a single load connected to the terminal of the faulted line. Although this approach is effective in conventional distribution systems, it is not applicable in distribution systems with installed DGs in downstream networks and systems with low short circuit capabilities in which fault currents are not significantly higher than the load currents. In this section, a method is proposed to address this problem. Depending on the availability of measuring devices in the downstream networks, two scenarios are considered:

*Scenario 1: No measuring device is installed downstream to the faulted line.*

In this case, an iterative procedure with the following steps should be followed.

Step 1: An initial value for the fault point $x_0$ is considered (e.g. $x_0 = 0.5$).

Step 2: Knowing the value of $x$, $v_F$ is calculated according to (1).

Step 3: Knowing the value of $v_F$, BFS method presented in Section II-B-2 is applied to the downstream network to calculate $v_{k+1}$ and $i_{k+1}$. The calculated value of $v_F$ in Step 2 is treated as the voltage of the first node in the explained BFS method in Section II-B-2 and the values of $v_{k+1}$ and $i_{k+1}$ are calculated accordingly.

Step 4: $i_F$ in Fig. 1-(b) is calculated as follows:

$$i_F = i_x - \frac{1}{2}xC\frac{dv_F}{dt} + i_{down} \qquad (61)$$

Where, $i_{down}$, the contribution of the downstream current to the fault current, is calculated as follows:

$$i_{down} = i'_x - \frac{1}{2}(1-x)C\frac{dv_F}{dt} \qquad (62)$$

$i'_x$ is calculated using (4) and

Step 5: Once $v_{k+1}$ and $i_{k+1}$ and $i_F$ are calculated in Step 4, the new value of $x$ is calculated according to Section II-A-4.

Step 6: if $|x^{iter} - x^{iter-1}| > \varepsilon$ or the number of iteration is less than the defined maximum number of iterations go to step 2.

Step 7: Stop and report Values of $i_{k+1}$, $v_{k+1}$, and $i_F$.

*Scenario 2: A measuring device is installed downstream to the faulted line.*

If there is at least one measuring device collecting synchronized samples downstream to the faulted line that measures both voltage and current signals, $v_{k+1}$ and $i_{k+1}$ can be calculated using similar procedure described in Section II-B. The node at which the measuring device collecting synchronized samples is installed is considered as the root-node in the explained procedure in Section II-B. To calculate $i_F$, an iterative procedure with the following steps should be followed.

Step 1: An initial value for the fault point $x$ is considered (e.g. $x_0 = 0.5$).

Step 2: Knowing the value of $x$, $v_F$ is calculated according to (1).

Step 3: $i_F$ in Fig. 1-(b) is calculated using (61)

Step 4: Once $v_{k+1}$ and $i_{k+1}$ and $i_F$ are calculated, the new value of x is calculated according to Section II-A-4.

Step 5: if $|x^{iter} - x^{iter-1}| > \varepsilon$ or the number of iteration is less than the defined maximum number of iterations go to step 2.

Step 6: Stop and report Values of $i_{k+1}$, $v_{k+1}$, and $i_F$.

| Algorithm 1: the procedure of the proposed fault location identification |
|---|
| 1: Start; |
| 2: $k \leftarrow 1$; |
| 3: Select line $k$; |
| 4: Calculate $v_k$ and $i_k$ as explained in Section II-B; |
| 5: Calculate $di_k/dt$ according to (47); |
| 6: Calculate $v_z$ according to (10); |
| 7: if scenario 1 holds then |
| 8:   Calculate $v_{k+1}$ and $i_{k+1}$ as explained in Section II-B; |
| 9: else |
| 10:   $iter \leftarrow 0$; |
| 11:   Assume $x^{iter} \leftarrow x_0$; |
| 12:   Calculate $v_F$ according to (1); |
| 13:   Calculate $v_{k+1}$ and $i_{k+1}$ using the proposed BFS method; |
| 14: end if |
| 15: Calculate $di_{k+1}/dt$ according to (47); |
| 16: Calculate $v'_z$ according to (12); |
| 17: Calculate $i_{down}$ according to (62); |
| 18: Calculate $i_F$ according to (61); |
| 19: Calculate $di_F/dt$ according to (47); |
| 20: Calculate $sg^+(i_F)$ and $sg^-(i_F)$ according to (6)-(7); |
| 21: $iter \leftarrow iter + 1$; |
| 22: Calculate $x^{iter}$ according to (34); |
| 23: if scenario 2 holds then |
| 24:   if $|x^{iter} - x^{iter-1}| > \varepsilon$ holds then |
| 25:     Go to Step 12; |
| 26:   end if |
| 27: end if |
| 28: if $0 < x < 1$ holds then |
| 29:   Add line $k$ to the short list; |
| 30: end if |
| 31: if $k < N_{bus} - 1$ holds then |
| 32:   $k \leftarrow k + 1$; |
| 33:   Go to Step 3; |
| 34: end if |
| 35: End |

It should be noted that in this case (i.e., measuring devices collecting synchronized samples are available at both upstream and downstream of the faulted line), there is an alternative approach for estimating the fault location. Instead of using (14), the following method can also be used. Similar to (8)-(9) the followings can be obtained:

$$v_k = xv_z + v_F \qquad (63)$$
$$v_{k+1} = (1-x)v'_z + v_F \qquad (64)$$

By subtracting (64) from (63) the following can be written:

$$v_k - v_{k+1} + v'_s = (v_s + v'_s)x \qquad (65)$$

Therefore, it can be solved using (34) as follows:

$$x = ([v_s + v'_s]^T W[v_s + v'_s])^{-1}[v_s + v'_s]^T W[v_k - v_{k+1} + v'_s] \quad (66)$$

### D. Estimating the Actual Location of the Fault Using Similarity Check

Once the short list of possible faulted lines is determined using the procedure presented in Section II-A, the actual location of the fault is determined using the following similarity check index. The index is based on the comparison between the measured values of current and/or voltages at the measuring devices' locations and the values of the same quantities (i.e., voltage and/or current values) calculated during the process of fault location identification based on the procedure explained in the previous subsections.

$$err = \frac{1}{N}\sum_{j=1}^{M}\sum_{i=1}^{N}(y_{ji}^m - y_{ji}^c)^2 \qquad (67)$$

Where M denotes the total number of measuring devices, $N$ denotes the total number of the collected data samples from each measuring device, superscripts $m$ and $c$ represent measured and calculated values, $y_{ji}$ represents ith data sample collected by jth measuring device respectively.

Note that in order to add up the differences in (67), they should be in per unit for current and voltage measurements. Consequently, after calculating differences in (67), for voltage and current measurements, the unit of differences is volts and Amperes, respectively. To represent the differences in per-unit, they are divided by the voltage and current base values, respectively.

Algorithm 1 summarizes the procedures of the proposed method for identifying the fault location.

## III. CASE STUDY RESULTS

The proposed time-domain based fault location identification method is applied to the modified IEEE 34-bus system, as shown in Fig. 1-(a). Two DGs are added to the distribution system at nodes 852 and 834. Also, lines on the main feeder from the substation to node 832 are replaced by tape-shielded 4/0 AWG aluminum underground cable. The impedance matrix of the cable is as follows [25]:

$$Z = \begin{bmatrix} 0.66 + j0.503 & j0.221 & j0.221 \\ j0.221 & 0.66 + j0.503 & j0.221 \\ j0.221 & j0.221 & 0.66 + j0.503 \end{bmatrix} \quad (68)$$

Also, the single-phase lateral connected to bus 816 is replaced by the same single-phase underground cable. In addition to the measuring device at the substation, two measuring devices are installed at buses 814 and 828 that can provide sampled values of the signals. The measuring device installed at node 814 is capable of measuring only voltage signal. However, the one installed at node 828 can measure both voltage and current signals. Thus, in the proposed algorithm, node 814 can never be considered a root node since its measurement is only voltage signal and can only be used in the similarity check process. However, node 828 can be considered a root node since it measures both voltage and current signals. Therefore, if the assumed faulted line is upstream to bus 828, the substation is considered as the upstream root done, and node 828 is considered as the downstream root node for the considered faulted line. In cases where the assumed faulted line is downstream to bus 828, the upstream root node of the line is node 828, and there is no downstream root node for that line. Moreover, there are current measurements at DG locations.

Starting from the substation, 10 different fault locations have been tested. The fault distance is increased in 3-mile intervals up to 30 miles. The voltage and current of a sub-cycle arc fault and an evolving fault 21 miles from the substation are shown in Fig. 4 and Fig 5, respectively. In the evolving fault, an A-G fault happens and after 3 cycles the fault evolves into an A-C-G fault. The results of the three different fault types, including A-G, A-B-G, and A-B are presented in Fig. 6, Fig. 7, and Fig. 8, respectively. Fault distance errors are calculated as follows:

$$Error = \frac{|Estimated\ Distance - Actual\ Distance|}{Actual\ Distance} \quad (69)$$

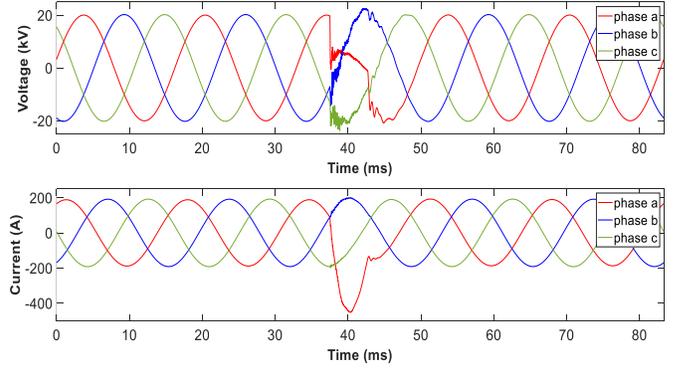

Fig. 4. Voltage and current of the faulted line during a sub-cycle fault

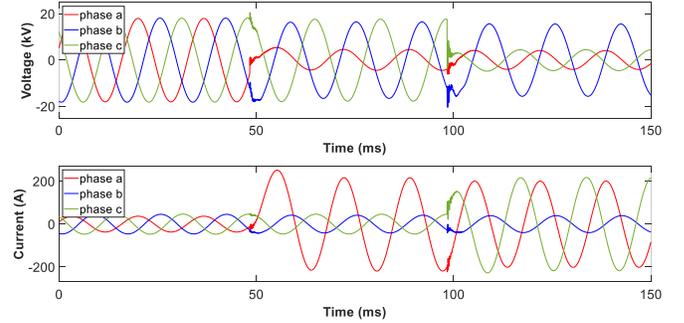

Fig. 5. Voltage and current of the faulted line during an evolving fault

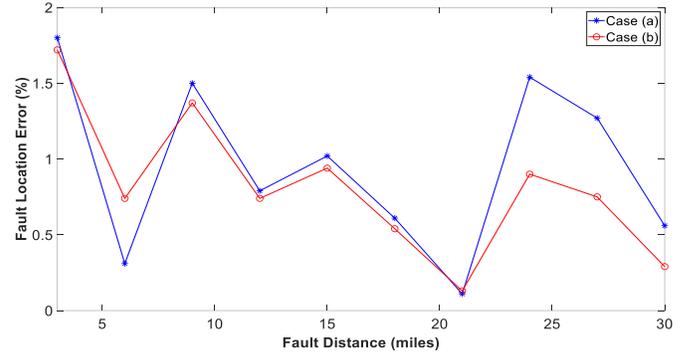

Fig. 6. Fault location error for an A-G fault, Case (a) $R = 10\ \Omega$, $L = 0.66315\ mH$, $v_{Fp} = 80\ V$, and $v_{Fn} = 100\ V$. Case (b) $R = 5\ \Omega$, $L = 0.66315\ mH$, $v_{Fp} = 80\ V$, and $v_{Fn} = 100\ V$.

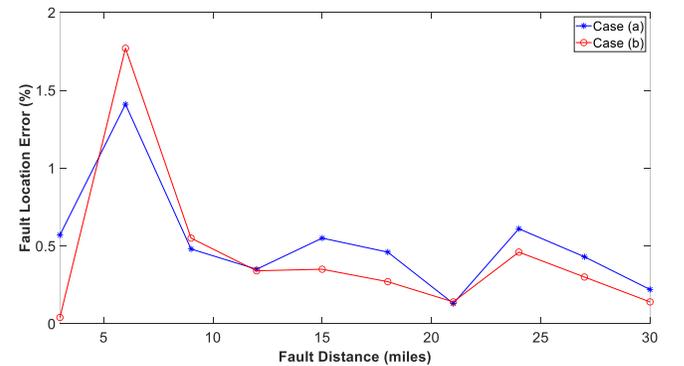

Fig. 7. Fault location error for an A-B-G fault, Case (a) $R = 10\ \Omega$, $L = 0.66315\ mH$, $v_{Fp} = 80\ V$, and $v_{Fn} = 100\ V$. Case (b) $R = 5\ \Omega$, $L = 0.66315\ mH$, $v_{Fp} = 80\ V$, and $v_{Fn} = 100\ V$.

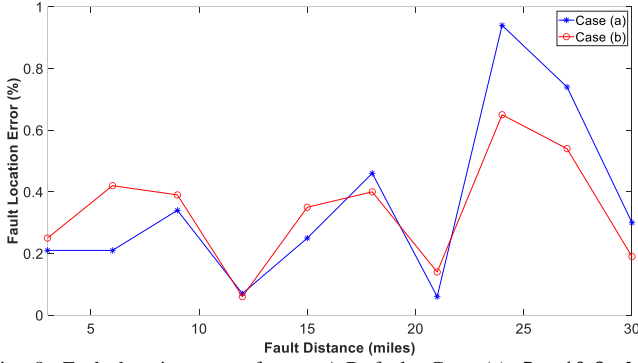

Fig. 8. Fault location error for an A-B fault, Case (a) $R = 10\ \Omega$, $L = 0.66315\ mH$, $v_{Fp} = 80\ V$, and $v_{Fn} = 100\ V$. Case (b) $R = 5\ \Omega$, $L = 0.66315\ mH$, $v_{Fp} = 80\ V$, and $v_{Fn} = 100\ V$.

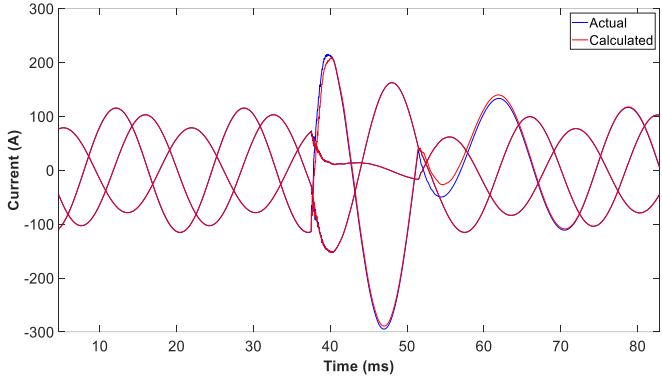

Fig. 9. Actual and calculated $i_{k+1}$ with load factor of 5 in the implemented time domain BFS method.

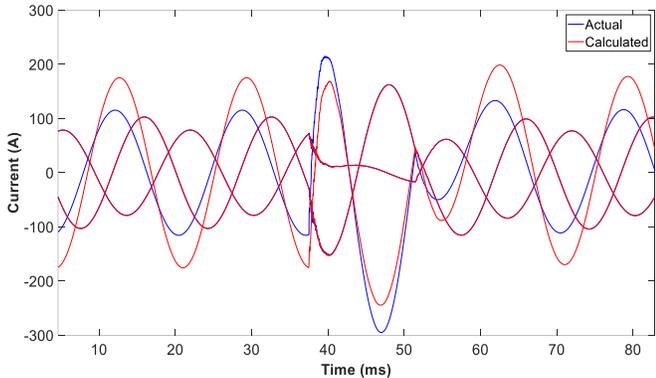

Fig. 10. Actual and calculated $i_{k+1}$ with the load factor of 5 when the lateral is replaced by the equivalent lumped up value of its loads.

Where distance denotes the electrical distance from the substation to the fault location. According to the case study results, all fault location errors are less than 2%.

To demonstrate the advantage of implemented time domain BFS method, a single-phase load is added to bus 822 with active and reactive powers of 500 kW and 300 kVAr, respectively. A single-phase A-G fault is considered 4 miles from the substation at the line between nodes 806 and 808. The active and reactive powers of the loads located at the lateral connected to bus 816 are multiplied by a factor with different values. Since this lateral is downstream to the faulted line, its load current affects $i_{k+1}$ and subsequently $v_{k+1}$. In the implement time-domain BFS

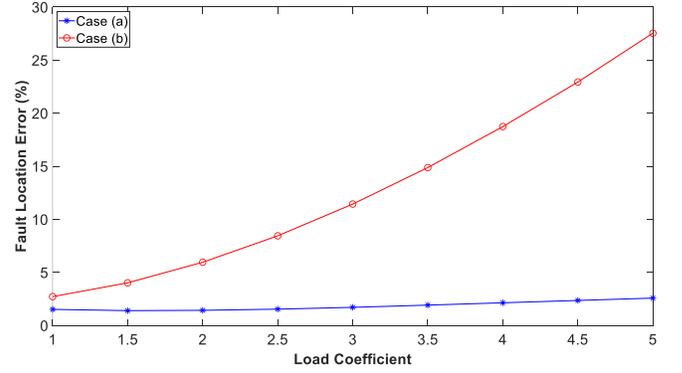

Fig. 11. Fault location error for different cases: Case (a) time domain BFS method is used, case (b): lateral is represented by the equivalent lumped up value of its loads.

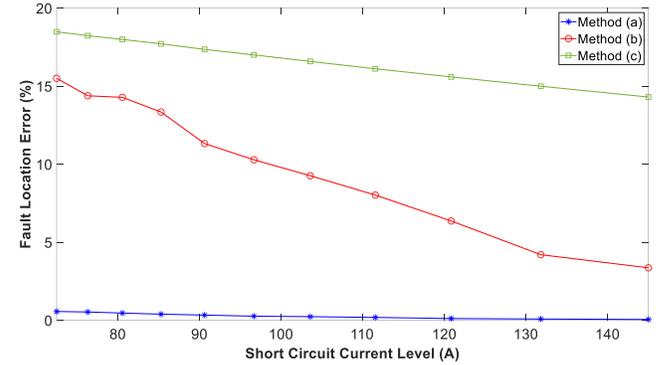

Fig. 12. Impacts of short circuit capacity of the system on different fault location methods: Method (a): The proposed method, Method (b): load currents are considered as measured pre-fault currents at the substation, Method (c) load currents are ignored.

method, each current of each load is calculated accurately, as shown in Fig. 9.

In Fig. 10, the implemented time-domain BFS is not used. Instead, all loads of the lateral are lumped up and located at bus 816. Fig. 10 represents the measured and calculated downstream current of the faulted line. As shown in Fig. 10, not using the implemented time-domain BFS method leads to a significant error in estimating the value of $i_{k+1}$. Moreover, Fig. 11 illustrates the fault location errors of both methods. As can be seen, when the time domain BFS method is used increasing the load values does not affect the fault location errors.

Fig. 12 shows the impacts of the short circuit capacity of the system on the proposed method in this paper and two other methods proposed in [23] and [27]. Note that [23] and [27] are developed for conventional distribution systems without DGs and cannot handle the presence of DGs. Therefore, in the reported results in Fig.12, DGs are removed from the system. In the proposed FL method in this paper, the load currents are calculated precisely, whereas in the literature, the load currents are either considered equal to pre-fault current measured currents at the substation [23] or the load currents are totally ignored [27]. Ignoring the load currents or assuming they are equal to the pre-fault values does not affect the results in small networks with high short circuit capacities and/or small loads. However, if loads of the distribution system are large and/or the short circuit capacity of the system is low, the load current must be calculated accurately. Fig. 12 illustrates fault location errors in three different methods. To decrease the short circuit capacity of the system, the source's impedance located at the substation is multiplied by a factor. It can be seen that the error of the proposed method remains low even when the short circuit capacity of the system is reduced.

## IV. CONCLUSION

In this paper, a time-domain fault location identification method was proposed. The proposed method can handle challenging fault cases such as sub-cycle faults and arc faults. Time domain backward/forward sweep method and load current calculation methods for fault location identification applications were proposed. The proposed fault location method can identify the actual location of the fault precisely without suffering from multiple fault location estimations. The proposed method is formulated in such a way that the impacts of networks downstream to the faulted line can be modeled accurately in the process of fault distance estimation. Different fault scenarios with different system short circuit capacities were studied. It was demonstrated that when load sizes increase and/or the short circuit capacity of the system reduces, the impacts of load currents become more influential. However, as the proposed method accounts for the presence of loads, laterals, and DGs, it is not affected by such impacts and is able to identify the fault location accurately.

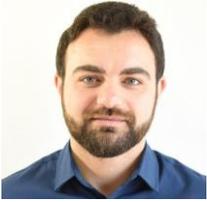

**Ali Shakeri Kahnamouei** (S'19) received his Ph.D. degree from Washington State University, Pullman, WA, USA. His research interests include power system protection and resilience.

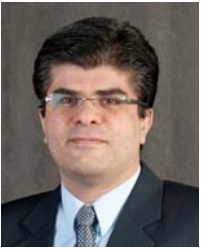

**Saeed Lotfifard** (S'08–M'11-SM'17) received his Ph.D. degree from Texas A&M University. Currently, he is an associate professor at Washington State University, Pullman. His research interests include stability, protection and control of inverter-based power systems. He serves as an associate editor for the IEEE Transactions on Power Delivery.